\documentclass[11pt]{article}
\usepackage{amssymb,amsmath}
\usepackage{amsfonts}
\usepackage{indentfirst}
\usepackage{nopageno}
\usepackage{graphicx}

\newcommand{\ds}{\displaystyle}
\title{ {\bf Stability bounds in a problem of convection with uniform internal heat source}}
\author{Ioana Dragomirescu$^{*}$, Adelina Georgescu$^{**}$\\
$^{*}$ Dept. of Mathematics, Univ. "Politehnica" of Timisoara, Romania,\\
 $^{**}$ Dept. of Mathematics, Univ. of Pitesti, Romania}

\begin{document}
\date{}

 \maketitle
 \pagestyle{empty}
\begin{abstract}
Motion in the atmosphere or mantle convection are two among phenomena of natural convection induced by internal
heat sources. They bifurcate from the conduction state as a result of its loss of stability.  In spite of their
importance, due to the  occurrence of variable coefficients in the nonlinear partial differential equations
governing the evolution of the perturbations around the basic equilibrium, so far these phenomena were treated
mostly numerically and experimentally. No rigorous study is known. In this paper we realize for the first time
such a linear study for the eigenvalue problem associated with those equations for a convection problem with an
uniform internal heat source in a horizontal fluid layer bounded by two rigid walls. Our method uses Fourier
series expansions for the unknown functions. Numerical results and graphs are given showing a destabilizing effect
of the presence of the heat source.
\end{abstract}
\section{Introduction} \par Problems of convection induced by an internal heat source have been experimentally (e.g. [5]) and numerically (e.g. [4]) studied by many researchers.  The investigations concerned the effects of the
heating and cooling rate, many investigations were performed for different conditions imposed on the lower and
upper boundaries. In these cases it is difficult to compare the evaluations of the heat fluxes in the heated fluid
layer  with the ones from the classical convection. The extent of the convection cells when the Rayleigh number
increases are greater than the ones in the classical Rayleigh-B\'{e}nard convection problem.
\par These problems are very important for the study of the motion in the atmosphere, influencing all other
processes that take place here.
\par In this paper a horizontal layer of viscous incompressible fluid with constant
viscosity and thermal conductivity coefficients $\nu$ and $k$ is considered [6]. In this context, the heat and
hydrostatic transfer equations [6] are
\begin{equation}
\label{eq:heat} \eta=k\dfrac{\partial^{2}\theta_{B}}{\partial z^{2}},
\end{equation}
\begin{equation}
\label{eq:hydro} \dfrac{dp_{B}}{dz}=-\rho_{B}g,
\end{equation}
where $\eta=const.$ is the heating rate, $\theta_{B}$, $p_{B}$ and $\rho_{B}$ are the potential temperature,
pressure and density in the basic state. In the fluid, the temperature at all point varies at the same rate as the
boundary temperature, so the problem is characterized by a constant potential temperature difference between the
lower and the upper boundaries $\Delta\theta_{B}=\theta_{B_{0}}-\theta_{B_{1}}$. Taking into account
(\ref{eq:heat}) this leads to the following formula for the potential temperature distribution [6]
\begin{equation}
\label{eq:potential_t}
\theta_{B}=\theta_{B_{0}}-\dfrac{\Delta\theta_{B}}{h}\Big(z+\dfrac{h}{2}\Big)+\dfrac{\eta}{2k}\Big[z^{2}-
\Big(\dfrac{h^{2}}{2}\Big)^{2}\Big].
\end{equation}
\par  In nondimensional variables the system of equations characterizing the problem is
\begin{equation}
\label{eq:ecuatie_nond} \left\{
\begin{array}{l}
\dfrac{d{\bf U}}{dt}= -\nabla p'+\Delta {\bf U}+Gr \theta' {\bf k},\\
\textrm{div} {\bf U}=0,\\
\dfrac{d\theta'}{dt}=(1-Nz){\bf U}{\bf k}+Pr^{-1}\Delta \theta',
\end{array}
\right.
\end{equation}
where ${\bf U}=(u,y,z)$ is the velocity, $\theta'$ and $p'$ are the temperature
 and pressure deviations from the basic state [6], $Gr$ is the Grashof number, $Pr$ is the Prandtl number and
 $N$ is a dimensionless parameter characterizing the heating (cooling) rate of the layer.
 \par The boundaries are considered rigid and ideal heat conducting, so the boundary conditions
 read
 \begin{equation}
 \label{eq:b_conditions}
 {\bf U}=\theta'=0 \textrm{ at } z=-\dfrac{1}{2} \textrm{ and } z=\dfrac{1}{2}.
 \end{equation}
 \vspace{0.2cm}
\section{Eigenvalue problem}
  \par Our analytical study has two major steps: we deduce the eigenvalue problem associated with
  (\ref{eq:ecuatie_nond})-(\ref{eq:b_conditions}) and then, we solve this problem using a modified method based on Fourier series expansions of
  Chandrasekhar [1] type.
  In [6] the numerical investigations concerned the vertical distribution of the total heat fluxes and their
 individual components for small and moderate supercritical Rayleigh number in the presence of a uniform heat
 source.
 \par In order to deduce the eigenvalue problem let us  consider the viscous incompressible fluid confined into a rectangular box bounded by two rigid walls:
  $V:0\leq x \leq a_{1}$, $ 0\leq y\leq a_{2}$, $-\dfrac{1}{2}\leq z\leq \dfrac{1}{2}$.
 First we deduce the eigenvalue problem governing the linear stability equivalent to this problem.
 We assume that any unknown function in (\ref{eq:ecuatie_nond}) is of the form from [3]
 \begin{equation}
 \label{eq:normal_modes}
 f(x,y,z)=\overline{F}(z)exp\Big(i\Big(\ds{2\pi m'\frac{x}{a_{1}}+2\pi n'\frac{y}{a_{2}}}\Big)\Big),
 \end{equation}
 $m=\dfrac{2\pi m'}{a_{1}},
 n=\dfrac{2\pi n'}{a_{2}}$, where $a_{1}=\dfrac{L}{H}$, $a_{2}=\dfrac{l}{H}$, $L$ and $l$ are the box sizes.
Here $m'\geq 1$ and $n'\geq 1$ are the number of cells in the $x$ and  the $y$ direction. Using
(\ref{eq:normal_modes}) we have
$
u=\overline{u}(z)e^{i(mx+ny)}, \ v=\overline{v}(z)e^{i(mx+ny)}, \ w=\overline{w}(z)e^{i(mx+ny)}, \
\theta'=\overline{\theta}(z)e^{i(mx+ny)}.
$
Applying to the equation $(\ref{eq:ecuatie_nond})_{1}$ the curl
operator, the system (\ref{eq:ecuatie_nond}) becomes
\begin{equation}
\label{eq:eigen_1} \left\{
\begin{array}{l}
(D^{3}\overline{v}-a^{2}D\overline{v})-in(D^{2}\overline{w}-a^{2}\overline{w})-in Gr \overline{\theta}=0,\\
D^{3}\overline{u}-a^{2}D\overline{u}-im(D^{2}\overline{w}-a^{2}\overline{w})-in Gr \overline{\theta}=0,\\
im(D^{2}-a^{2})\overline{v}-in(D^{2}-a^{2})\overline{u}=0,\\
(D^{2}-a^{2})\overline{\theta}Pr^{-1}+(1-Nz)\overline{w}=0,\\
im\overline{u}+in\overline{v}+D\overline{w}=0,
\end{array}
\right.
\end{equation}
 where $a^{2}=m^{2}+n^{2}$ and the boundary conditions are
\begin{equation}
\label{eq:eigen_bc1} \overline{u}=\overline{v}=\overline{w}=\overline{\theta}=0 \textrm{ at } z=\pm \dfrac{1}{2}.
\end{equation}
From $(\ref{eq:eigen_1})_{3}$, taking into account the boundary conditions (\ref{eq:eigen_bc1}), we get
$\overline{u}=\dfrac{m}{n}\overline{v}$. Replacing this expression in $(\ref{eq:eigen_1})_{5}$, we obtain
$\overline{v}=\dfrac{in}{a^{2}}D\overline{w}$. In this way, the system (\ref{eq:eigen_1}) and the boundary
conditions (\ref{eq:eigen_bc1}) reduce to the following two-point problem
\begin{equation}
\label{eq:eigen_2} \left\{
\begin{array}{l}
(D^{2}-a^{2})^{2}\overline{w}-a^{2}Gr \overline{\theta}=0,\\
(D^{2}-a^{2})\overline{\theta}+Pr(1-Nz)\overline{w}=0,
\end{array}
\right.
\end{equation}
\begin{equation}
\label{eq:eigen_bc2} \overline{w}=D\overline{w}=\overline{\theta}=0 \textrm{ at } z=-\dfrac{1}{2} \textrm{ and }
z=\dfrac{1}{2}.
\end{equation}
\section{Secular equation}
\par Taking into account the boundary conditions imposed for the normal component of the velocity and for the normal
component of the derivative of the velocity, a method based on Fourier series expansions upon complete sets of
orthogonal trigonometric functions that satisfy all boundary conditions it is not possible to apply. In this way,
at first, we expanded the unknown function $\overline{w}$ upon a complete set of orthogonal Chandrasekhar type
functions (see Chandrasekhar [1]), but here these functions introduced an extraperiodicity to the problem leading
to the lose of one of the physical parameters, e.g. the heating (cooling) rate $N$.
\par In order to avoid this problem, let us modify the system (\ref{eq:eigen_2}) by a translation of the variable $z$,
$x=z+\dfrac{1}{2}$, such that the eigenvalue problem becomes
\begin{equation}
\label{eq:eigen_3} \left\{
\begin{array}{l}
(D^{2}-a^{2})^{2}\overline{w}=a^{2}Gr \overline{\theta},\\
(D^{2}-a^{2})\overline{\theta}=-Pr(N_{1}-Nx)\overline{w},
\end{array}
\right.
\end{equation}
where $N_{1}=1+\dfrac{N}{2}$, with the boundary conditions
\begin{equation}
\label{eq:eigen_bc3} \overline{w}=D\overline{w}=\overline{\theta}=0 \textrm{ at }x=0 \textrm{ and }1.
\end{equation}
\par Eliminating the unknown function $\overline{w}$ between the two equations from (\ref{eq:eigen_3}),
 the following differential equation in the unknown function $\overline{\theta}$  only
\begin{equation}
\label{eq:ecuatie_in_t}
(D^{2}-a^{2})^{3}\overline{\theta}+Ra(N_{1}-Nx)\overline{\theta}=0
\end{equation}
is obtained, with $ \overline{\theta}=(D^{2}-a^{2})\overline{\theta}=D(D^{2}-a^{2})\overline{\theta}=0 \textrm{ at
} x=0 \textrm{ and }1.
$
\par Using the same method described in detail in [1], we introduce a new function $\Psi$ and rewrite the equation
(\ref{eq:ecuatie_in_t}) as a new system in the unknown functions $\overline{\theta}$ and $\Psi$
\begin{equation}
\label{eq:sistem_simplu} \left\{
\begin{array}{l}
(D^{2}-a^{2})^{3}\overline{\theta}=(N_{1}-Nx)\Psi,\\
\Psi=-a^{2}Ra \overline{\theta},
\end{array}
\right.
\end{equation}
where $Ra=Gr\cdot Pr$ is the Rayleigh number. The Rayleigh number is an eigenvalue in (\ref{eq:sistem_simplu}) and
$(\overline{\theta}, \Psi)$ represent the corresponding eigenvector. We are interested in finding the smallest
eigenvalue in (\ref{eq:sistem_simplu}), defining the neutral manifold. Taking into account the boundary conditions
for $\overline{\theta}$, we set $ \Psi=\sum\limits_{m=1}^{\infty}\Psi_{k}\sin k\pi x,$ \ \
$\overline{\theta}=\sum_{k=1}^{\infty}\Psi_{k}\Theta_{k}, $ where $\Theta_{k}$ is found from the equation
\begin{equation}
\label{eq:ecuatie_neomogena} (D^{2}-a^{2})^{3}\Theta_{k}=(N_{1}-Nx)\sin k\pi x.
\end{equation}
Equation (\ref{eq:ecuatie_neomogena}) has the solution
\begin{equation}
\label{eq:eq15} \begin{array}{l}
\Theta_{k}=-\dfrac{1}{(k^{2}\pi^{2}+a^{2})^{3}}\Big[(A_{0}^{k}+A_{1}^{k}x+A_{2}^{k}x^{2})\cosh
ax+(B_{0}^{k}+B_{1}^{k}x+B_{2}^{k}x^{2})\sinh ax+\\
+(N_{1}-Nx)\sin k\pi x-\dfrac{6k\pi N}{(k^{2}\pi^{2}+a^{2})}\cos k\pi x\Big],
\end{array}
\end{equation}
The unknown coefficients $A_{0}^{k}$, $A_{1}^{k}$, $A_{2}^{k}$, $B_{0}^{k}$, $B_{1}^{k}$, $B_{2}^{k}$ are obtained
 from the system followed by imposing to the function $\Theta_{k}$ to satisfy the boundary conditions imposed on
$\overline{\theta}$
\begin{equation}
\label{eq:eq16} \left\{
\begin{array}{l}
A_{0}^{k}=\dfrac{6k\pi N}{k^{2}\pi^{2}+a^{2}},\\
\\
(A_{0}^{k}+A_{1}^{k}+A_{2}^{k})\cosh a+(B_{0}^{k}+B_{1}^{k}+B_{2}^{k})\sinh a=\dfrac{6k\pi
N }{k^{2}\pi^{2}+a^{2}}(-1)^{k},\\
\\
A_{2}^{k}+aB_{1}^{k}=-2k \pi N,\\
\\
a\sinh aA_{1}^{k}+A_{2}^{k}(\cosh a+2a\sinh a)+a\cosh aB_{1}^{k}+B_{2}^{k}(\sinh a+2a\cosh a)=\\
=2k \pi N(-1)^{k+1},\\
\\
a^{2}A_{1}^{k}+3aB_{2}^{k}=\dfrac{N_{1}}{2}k\pi(k^{2}\pi^{2}+a^{2}),\\
\\
a^{2}\cosh aA_{1}^{k}+A_{2}^{k}(2a^{2}\cosh a+3a\sinh a)+a^{2}\sinh aB_{1}^{k}+B_{2}^{k}(3a\cosh a+2a^{2}\sinh
a)=\\
=\dfrac{N_{1}-N}{2}k\pi (-1)^{k}(k^{2}\pi^{2}+a^{2}).
\end{array}
\right.
\end{equation}
Replacing the obtained expressions for the coefficients $A_{i}^{k}$, $B_{i}^{k}$, $i=0,1,2$ in the equation
$(\ref{eq:sistem_simplu})_{2}$  and imposing the condition that the obtained equation  be orthogonal to the
functions $sin l\pi z$, $l\in \mathbb{N}$ with respect to the inner product of $L^{2}(0,1)$,
$(f,g)=\ds\int_{0}^{1}fgdz$, we obtain an infinite system of algebraic linear equations in $\Psi_{k}$.
\par The secular equation is obtained by imposing the condition that the determinant of this
system to vanish, i.e.
\begin{equation}
\label{eq:eq17}
\begin{array}{l}
det
\Big\|\dfrac{\delta_{kl}}{2}=\dfrac{a^{2}Ra}{(k^{2}\pi^{2}+a^{2})^{3}}[A_{0}^{k}I^{l}_{01}+A_{1}^{k}I^{l}_{11}+A_{2}^{k}I^{l}_{21}
+B_{0}^{k}I^{l}_{02}+B_{1}^{k}I^{l}_{12}+B_{2}^{k}I^{l}_{22}+N_{1}\dfrac{\delta_{kl}}{2}-\\
\\
-NT_{kl}+\dfrac{6k\pi N}{k^{2}\pi^{2}+a^{2}}U_{kl}]\Big\|,
\end{array}
\end{equation}
where  $T_{kl}=\left\{
\begin{array}{l}
\ds \frac{1}{4} \textrm{ if }k=l\\
\\
\ds \frac{2kl[(-1)^{k+l}-1]}{\pi^{2}(l-k)^{2}(l+k)^{2}}\ \ \  \textrm{ if }k\neq l
\end{array},
\right.$ $U_{kl}=\left\{
\begin{array}{l}
0 \textrm{ if }k=l\\
\\
\ds \frac{l[(-1)^{k+l}-1]}{\pi(l-k)(l+k)} \textrm{ if }k\neq l
\end{array}
\right.$, $$I^{l}_{ij}=\left\{ \begin{array}{l} \ds\int_{0}^{1}x^{i}\cosh(ax)\sin l\pi x dx,  \textrm{ if } j=1\\
\ds\int_{0}^{1}x^{i}\sinh(ax)\sin l\pi x dx, \textrm{ if } j=2, i=0,1,2.
\end{array}
\right.$$ From the secular equation, keeping some of the parameters fixed, we obtain the neutral manifolds, e.g
neutral curves or neutral surfaces. \section{Numerical results} In this case, a first approximation for the
Rayleigh number $Ra$ is obtained by taking $k=l=1$ and imposing to the first-order minor in the matrix associated
with the infinite algebraic system in $\Psi_{k}$ to vanish [1]. But for $k=l=1$ the expression of the Rayleigh
number following from the secular equation is  given by $$
Ra=\dfrac{(\pi^{2}+a^{2})^{5}(a+\sinh(a))}{a^{2}[(\sinh(a)+a)(a^{2}+\pi^{2})^{2}-8a \pi^2(1+\cosh(a))]}.
$$ In this expression the parameter $N$ does not occur so, we cannot use this approximation. We performed
numerical evaluations for the Rayleigh number for the second and the third approximation, i.e. $k=l=2$ and $k=l=3$
respectively. The numerical results are presented in Table 1. \par From the approximate numerical evaluations it
can be seen that when the wavenumber $a$ is increasing, the approximate Rayleigh number is also increasing. When
the wavenumber is kept constant an increase in the heating (cooling) rate parameter leads to a decreasing of the
Rayleigh number. Moreover, for values of the parameter $N$ greater than a certain value (stabilized around 10),
the third approximation is greater, yet not very different, from the one obtained in the second approximation.
\par In fig.1 the approximate neutral curve for the parameter characterizing the heating rate $N=1$ is presented.
When both parameters are varying the approximate neutral surface
is obtained (fig.2).
\begin{center}
\begin{tabular}{|c|c|c|c|c|}
\hline $N$&$a^{2}$&$ Ra (k=l=2)$ &$Ra (k=l=3)$&$Ra (var. meth.)$\\ \hline $0$&$9.711$&$1715.079324$&$1715.079324$&$1749.97575$\\
\hline $1$&$9.711$&$1711.742588$&$1704.733019$&$1746.80944$\\
\hline $2$&$9.711$&$1701.891001$&$1695.265991$&$1737.45025$\\
\hline $1$&$10.0$&$1712.257687$&$1705.203119$&$1747.29100$\\
\hline $4$ &$10.0$&$1664.341789$&$1659.087870$&$1701.62704$\\
\hline $4$&$12.0$&$1685.422373$&$1680.142966$&$1723.62407$\\
\hline $8$&$12.0$&$1547.460446$&$1546.437526$&$1590.19681$\\
\hline $9$&$12.0$&$1508.147637$&$1508.070640$&$1551.72378$\\
\hline $10$&$12.0$&$1468.449223$&$1469.223193$&$1512.69203$\\
\hline $12$ &$12$&$1389.837162$&$1392.166660$&$1434.90396$\\
\hline $16$&$12$&$1243.442054$&$1247.400439$&$1288.50149$\\
\hline $10$&$9.0$&$1482.527042$&$1482.391699$&$1525.59302$\\
\hline $11$&$9.0$&$1446.915467$&$1447.488638$&$1490.55802$\\
\hline $12$&$9.00$&$1411.401914$& $1412.610226$&$1455.48233$\\
\hline $30$&$12.00$&$879.104231$&$884.196861$&$917.507873$\\
\hline $50$&$9.00$&$643.4478727$&$647.561188$&$673.848081$\\
\hline
\end{tabular}
\end{center}
\begin{center}
{\bf Table 1 } Numerical evaluations of the Rayleigh number for various values of the parameters $N$ and $a$.
\end{center}
\begin{figure}[h]
\begin{center} \includegraphics[height=4.5cm,width=5cm]{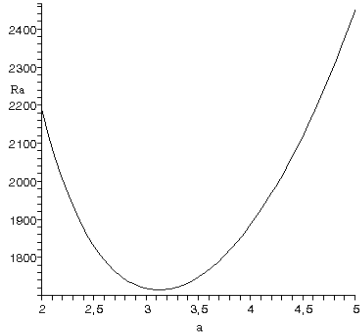}
\end{center}
\caption{The approximate neutral surface $Ra=Ra(a)$ for $N=1$.} \label{1}
\end{figure}
\par We also verified the numerical results with a variational method: the Rayleigh quotient method.
The obtained numerical results are also presented in Table 1. However the approximations are limited by the
difficult evaluation of the associated matrix for a large number of functions in the expansion sets. Also, in this
cases, the convergence it is not always assured. The advantage in using this method is that the expressions of the
neutral manifolds are easy to obtain.

\begin{figure}[h]
\begin{center}
\includegraphics[height=4.5cm,width=5cm]{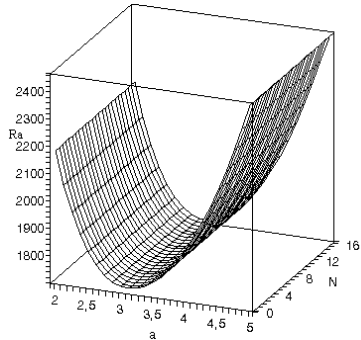}
\end{center}
\caption{The approximate neutral surface $Ra=Ra(a,N)$.} \label{2}
\end{figure}
\par When $N=0$ the problem reduces to the particular case of Rayleigh-B\'{e}nard convection and the numerical
evaluation lead us to a critical value similar to the classical value for the Rayleigh number, i.e.
$Ra=1715.079324$ for $a=3.117$. \section{Conclusions} \par In this paper we are concerned with the stability of
the mechanical equilibrium of a fluid layer, namely a problem of convection with uniform internal heat source in a
parallelipipedic box. The eigenvalue problem is obtained and then investigated in order to obtain the secular
equation which gives the approximate stability limits. \par Numerical investigations were performed in order to
establish the influence of the two parameters, the wavenumber and the parameter characterizing the heating
(cooling) rate on the approximate values of the Rayleigh number: when the heating (cooling) rate is increasing,
the domain of stability is decreasing and an increase in the wavenumber enlarge the domain of stability. We
pointed out that the first approximation does not work in this case and remarked that some other classical method
do not work either in this case. \end{document}